\documentclass[twocolumn]{oja}
\usepackage{lmodern}
\usepackage{amsmath}
\usepackage{xcolor} 
\usepackage{hyperref}
\usepackage{orcidlink}

\renewcommand\orcidlink[1]{\href{https://orcid.org/#1}{\includegraphics[width=8pt]{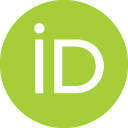}}}
\def\figwidsolo{3.475in}
\def\figwiddbl{5.75in}

\def\CMD{HR diagram}  %

\def\AU{{\rm au}}
\def\Msolar{\ifmmode {\rm M_{\odot}}\else $\rm M_{\odot}$\fi}
\def\systar{\texttt{Sy*}}
\def\systarmaybe{\texttt{Sy?}}

\newcommand{\simbadvar}[1]{\texttt{#1}} 
\newcommand{\gaiavar}[1]{\texttt{#1}} 
\def\ruwe{\gaiavar{ruwe}}
\def\sourceid{\gaiavar{source\_id}}
\def\xpcontinuous{\gaiavar{has\_xp\_continuous}}
\def\photgmeanmag{\gaiavar{phot\_g\_mean\_mag}}
\def\bprp{\gaiavar{bp\_rp}}
\def\absmag{$M_G$}
\def\ra{\gaiavar{ra}}
\def\dec{\gaiavar{dec}}
\def\parallax{\gaiavar{parallax}}
\def\poe{\gaiavar{parallax\_over\_error}}
\def\agof{\gaiavar{astrometric\_gof\_al}}

\def\pmra{\gaiavar{pmra}}
\def\pmdec{\gaiavar{pmdec}}
\def\projco{S} 
\def\cof{$proj\_coeff$}

\def\rosat{{\it ROSAT}}

\def\xmm{{\it XMM--Newton}}
\def\erosita{{\it SRG/eROSITA}}
\def\2mass{{\it 2MASS}}

\def\gaia{{Gaia}}
\def\GAIA{{GAIA}}

\def\Halpha{H$\alpha$}
\def\Hbeta{H$\beta$}
\def\Hgamma{H$\gamma$}
\begin{document}

\title{Symbiotic star candidates in GAIA Data Release 3}

\shorttitle{Symbiotic star candidates}

\author{Samantha E. Ball$^1$\orcidlink{0009-0004-1752-2953}}   
\author{Benjamin C. Bromley$^1$\orcidlink{0000-0001-7558-343X}}
\author{Scott J. Kenyon$^2$\orcidlink{0000-0003-0214-609X}} 

\affiliation{$^1$Department of Physics and Astronomy, University of Utah, 115 S 1400 E, Salt Lake City, UT, 84112, USA}
\affiliation{$^2$Smithsonian Astrophysical Observatory, 60 Garden Street, Cambridge, MA 02138, USA} 
\email{samantha.e.ball.00@gmail.com} 
\email{bromley@physics.utah.edu}
\email{skenyon@cfa.harvard.edu}

\shortauthors{Ball, Bromley, and Kenyon}

\begin{abstract}
Symbiotic stars, binary pairs with a cool giant fueling accretion onto a hot compact companion, offer unique insights to our understanding of stellar evolution. Yet, only a few hundred symbiotic stars are confirmed. Here, we report on a new search for symbiotic star candidates in \gaia\ Data Release 3 (GDR3), based entirely on the archive's astrometric, photometric, and spectroscopic information. To begin our search, we identified known symbiotic stars in GDR3 and assessed their absolute magnitude and colors, which are dominated by the cool giant. We also considered measures of astrometric quality that might be affected by binary motion in these systems. Finally, from those sources with \gaia\ spectroscopic data, we built a low-resolution spectral template that characterizes the unique features of these systems, including \Halpha\ emission from interaction with the giant's wind and radiation from the hot star. We then queried the full GDR3 archive for sources with spectroscopic data that are bright ($<$~17 mag in G-band), have modest relative parallax uncertainties ($<$~20\%), and fall within a region of color–magnitude space characteristic of red giants, keeping only sources with spectra that quantitatively match our template. A machine-learning algorithm, trained on known symbiotic stars, produced a new catalog of 1,674 sources. From cross-matches with infrared and X-ray surveys, we present 25 of these sources as particularly compelling candidates for new symbiotic stars. 
\end{abstract}

\keywords{Symbiotic stars, Machine learning, Spectroscopy}

\section{Introduction} 
\label{sec:intro}

Symbiotic stars are cosmic rarities. They are interacting binaries with a coolgiant star shedding mass onto a hot compact companion \citep{boyarchuk1969, bath1977, allen1984, kenyon1984,  kenyon1986, webbink1987, munari2019, merc2025review}.  The hotter star --- usually a white dwarf --- accretes material from a wind or Roche-lobe overflow from its cooler companion, a red giant branch (RGB) or asymptotic giant branch (AGB) star. When the hot star is a white dwarf, mass transfer generates additional energy from nuclear burning, driving occasional outbursts or steady high-energy emission \citep[e.g.,][]{payne1964,allen1980,kenyon1983,mikola1992,luna2013,munari2025}. A more spectacular outcome, a Type Ia supernova, may result if the compact star reaches its Chandrasekhar limit \citep{kenyon1993sysupernovae, hachisu1999, liu2023}. Thus, as progenitors of the premier standard candle in cosmology \citep{riess1998}, symbiotic stars play a critical role in our understanding of the overall evolution of the universe.  

The unusual configuration of symbiotic stars, each with a hot and cool binary partner, leads to a unique spectral signature. Thermal emission from the compact hot star contributes to a blue continuum, while the cooler giant dominates the red spectrum. Molecular absorption lines from the giant star's atmosphere appear alongside emission from ionized species including \mbox{[O\,III]}, He\,II, and hydrogen (\Halpha, \Hbeta) \citep{kenyon1987}, as the compact star ionizes material lost by the giant.  This combination of a double-peaked continuum, absorption bands and emission lines from highly ionized atoms distinguishes symbiotic stars from other interacting binary systems, including cataclysmic variables (a white dwarf and main-sequence companion) and X-ray binaries (a main sequence star in close orbit with a neutron star or black hole). These same features --- evidence of a cool giant with molecular absorption features coupled with the clear signature of hot gas --- guided pioneers in the discovery of symbiotic stars \citep{merrill1932, swings1941, paynegaposchkin1946}. 

Other key observational features of symbiotic stars emerge in photometric imaging and astrometric data. Photometry captures brightness variations that enable precise determinations of stellar magnitudes, temperatures, and variability. In symbiotic stars, the complex interactions between binary partners lead to distinctive light curves, including periodic and eruptive variations \citep[e.g.,][]{merc2020}. Symbiotic binary partners also tend to have long orbital periods compared with other interacting binaries, ranging from hundreds of days (for S-type stars) to decades \citep[D-type systems, e.g.,][]{gromadzki2013}. The corresponding orbital separations are typically $\gtrsim$ 2--3~au, enabling high-resolution imaging to distinguish components \citep[e.g.,][]{schmid2017, bujarrabal2018}. Astrometric data from symbiotic stars contributes to refinements of stellar luminosities as well as their distribution within the Galaxy. These unique characteristics allow symbiotic systems to be distinguished from other binary and single-star systems. 

To date, over 300 symbiotic stars are listed in the SIMBAD astronomical database. This census, while impressive given observational challenges in the identification of individual stars \citep[e.g.][]{merc2025review}, is significantly less than theoretical predictions for the population of symbiotic stars in the Milky Way, which range from 10$^3$ to 10$^5$ \citep{kenyon1986, mmunari1992, magrini2003, lu2006, laversveiler2025}. This discrepancy continues to drive searches for these rare objects, with strategies that focus on optical line emission surveys \citep{munari2021}, along with photometry in optical \citep{lucy2024}, IR \citep{akras2019a, akras2021, chen2025}, UV \citep{akras2023}, and X-ray \citep{xu2024, merc2025xray, guerrero2025} wave bands. These studies, including others that incorporate machine learning \citep{akras2019b, jia2023}, have yielded dozens of candidates, though far fewer confirmed symbiotic stars.

Here our focus is on the discovery of symbiotic stars, casting a wide, loosely knit net to retrieve high-potential sources for follow-up observations. We use observations from the European Space Agency's \gaia\ satellite, which has provided astrometry and photometric data for over a billion stars \citep[Data Release 3, hereafter GDR3]{gaiadr32023}. Gaia data have guided previous searches for symbiotic stars as part of broader classifications of variable stars \citep{eyer2023, rimoldini2023}. However, pulsating stars are a significant source of contamination \citep[e.g.,][]{merc2025review}.Our approach takes advantage of Gaia's low-resolution spectra available for over 200~million sources.  Using known symbiotic stars from SIMBAD \citep{simbad2000}, we train a machine learning (ML) algorithm to identify promising candidates \citep{akras2019b, jia2023}. We describe this approach in \S\ref{sec:method}. Then we apply the ML algorithm to a subset of stars in GDR3 to obtain a new catalog of these candidates (\S\ref{sec:results}). We conclude in \S\ref{sec:conclude}.

\section{A catalog of known symbiotic stars} 
\label{sec:cat}

Symbiotic stars have emerged as a distinct class of stars because of their spectral features. Indeed, these objects began to stand out roughly a century ago as a result of discoveries by Cannon, Merrill and Payne-Gaposhkin \citep[see][for a historical summary]{kenyon1986}. While \gaia\ is foremost an astrometric database, it provides photometry, even in the time domain \citep[cf.][]{merc2020}. It also contains spectroscopic information in the form of spectral template coefficients for about 10\%\ of the DR3 sources. Our approach here is to gather this information for known symbiotic stars and to train a machine learning algorithm to identify promising symbiotic star candidates. We begin with our collection of known symbiotic stars.

\subsection{Symbiotic stars in \gaia\ with low-resolution spectra}

To identify known symbiotic stars, we queried the SIMBAD database \citep{simbad2000} seeking sources with object types \systar\ and skipping stars listed as object type \systarmaybe. The result yields 302 known symbiotic stars. SIMBAD provided a GDR3 \sourceid\ for 243 of these objects. We reviewed the leftover 59 stars with Aladin Lite \citep{aladinlite} using the coordinates given in SIMBAD to manually match GDR3 sources. Of the 59 stars, five have both a manually matched GDR3 \sourceid\ and \gaia\ spectral data (true-valued \xpcontinuous\ flag). One star, with SIMBAD \simbadvar{main\_id} EM*~AS~269, was excluded because there were two GDR3 close sources with comparable magnitudes, making it difficult to distinguish the symbiotic star. This procedure defines a sample set of 248 sources that are recognized in SIMBAD as symbiotic stars, with 221 objects that have GDR3 spectra. Table~\ref{tab:sysstars} below lists the five objects we matched to GDR3, while the online version has all 248 stars. Of these sources, 224 are listed as confirmed symbiotic stars in Table 6 of \citet{akras2019a}. Similarly, an actively managed repository, the New On-line Database of Symbiotic Variables \citep[NODSV]{merc2019}, contains 234 of our sources, with 214 designated as confirmed, and 20 listed as at least possible symbiotic stars.

\setlength{\tabcolsep}{4pt}
\begin{deluxetable}{lllr}
\tablecaption{Symbiotic stars manually matched with spectra data
\label{tab:sysstars}
}
\tabletypesize{\scriptsize}
\tablehead{ \colhead{SIMBAD} & \colhead{ra (ICRS)} & \colhead{dec (ICRS)} & \colhead{\gaia\ \sourceid}}
\startdata
V* V916 Sco   & 17 43 54.7 & $-$36 03 26.5 &  4040972692694908928 \\
PN K  3-22  & 19 09 26.6  &  +12 00 44.6 &   4313125288215049728 \\
V* V4073 Sgr &  18 08 19.1  & $-$22 57 19.0 &  4066601273532526336 \\
PN Th  3-29   & 17 32 27.9 & $-$29 05 08.8&   4058843158219588480  \\
VVV NV003  & 17 50 19.3 & $-$33 39 07.3 &   4041878930741533824
\enddata
\end{deluxetable}
\setlength{\tabcolsep}{6pt}

\subsection{Photometric properties}\label{subsec:phot}

\begin{figure}
    \centering
    \includegraphics[width=\figwidsolo]{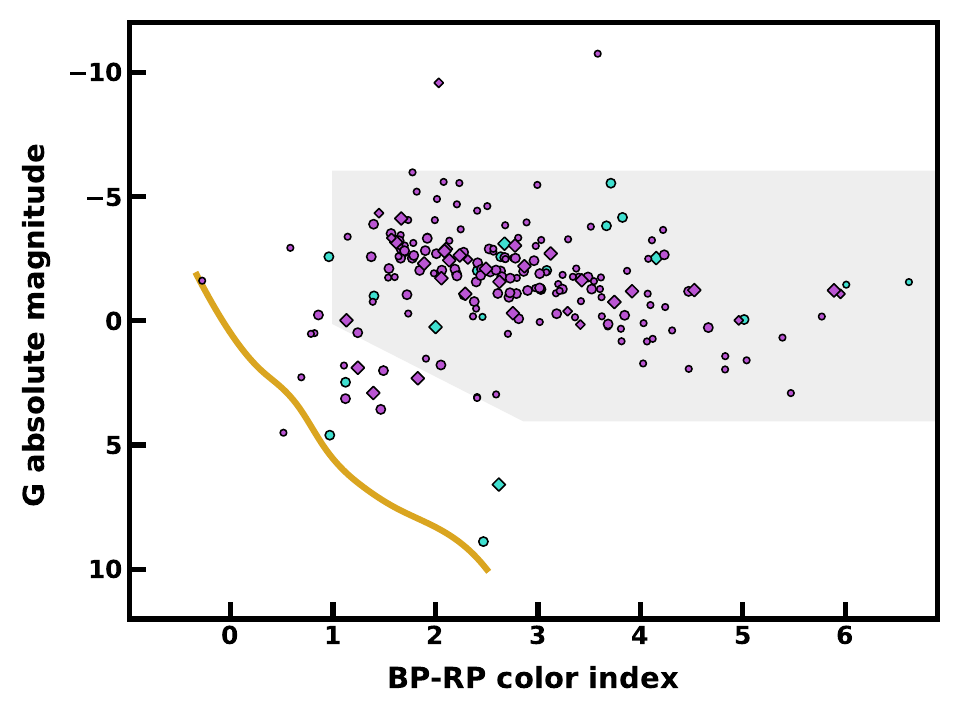}
\caption{A Hertzsprung-Russel diagram for the 248 symbiotic stars highlighted in this study. Each point is derived from GRD3 parameters. Most of the sources (224, violet markers) are confirmed in \citet[][Tables 6+7 therein]{akras2019a}, while the remaining 24 were identified here with SIMBAD (blue-green markers). The larger points indicate that the GDR3's parallax error is less than 20\%. The shape of the symbol indicates whether a star is associated with an X-ray source (diamond) or not (circle).The smaller points thus correspond to sources that have less certain distance moduli. For reference, a cartoon curve (lower left) illustrates the approximate location of the main-sequence ridge in the \gaia\ survey \citet[e.g.,][also, \href{https://sci.esa.int/s/wKdy4dW}{https://sci.esa.int/s/wKdy4dW} with Gaia's \CMD]{bailerjones2021}. The plot also includes a black dot and tail, showing the displacement of a hot blue star in the color-magnitude plane from reddening of 1~mag. The extinction is derived from the extinction-to-color excess ratio $R = 3.25$ \citep{li2023, rzhang2023}.\label{fig:systs_cmd}}
\end{figure}

We illustrate attributes of the 248 stars selected here, starting with a Hertzsprung-Russel (HR) diagram in Figure~\ref{fig:systs_cmd}. The \CMD\ emphasizes that the cool components of symbiotic systems are post-main-sequence RGB or AGB stars, with sources broadly clustered around a \bprp\ color index of +2 and an absolute magnitude of around –2 in Gaia’s G band. There are a few exceptions, with a handful of sources that appear near the main sequence, and two that are unphysically bright. These outliers tend to have poorly constrained distances (low \poe) or high levels of extinction (e.g., high \gaiavar{ag\_gspphot}).  

 Figure~\ref{fig:systs_cmd} also indicates potential cross-matches in all-sky X-ray surveys from \rosat\ \citep{aschenbach1981, white2000}, \xmm\ \citep{jansen2001, webb2020}, and \erosita\ \citep{predehl2006, freund2024}, with a total of 42 sources identified in at least one of these catalogs (Table~\ref{tab:sysstars}, online version). We made these identifications with positional searches using search radii of 30~arcsec, 5~arcsec, and 20~arcsec for \rosat, \xmm, and \erosita, respectively. As a check on this approach, 32 of our X-ray cross-matches are reported in the NODSV as X-ray detections. Two of our cross-listings are designated as having no detected X-ray emission; the remaining eight stars are neither confirmed nor refuted as X-ray emitters.

Almost a quarter of our selected stars are plausible X-ray emitters, which is a much larger fraction than in the general population of red giants \citep[e.g.,][]{hunsch1998, schmitt2024} as X-ray production in symbiotic systems stems from interactions between their binary components. In particular, X-ray emission can originate from a hard, highly absorbed component likely produced in an accretion-disk boundary layer, a soft component consistent with shocks from wind-wind collisions, or a combination of both mechanisms, with supersoft emission in some systems attributed to nuclear shell burning on the white dwarf \citep[e.g.,][]{luna2013}. In contrast, such interactions are absent in single red giants, making comparable X-ray emission rare. 

Figure~\ref{fig:systs_cmd} shows a ``selection zone'' in the color-magnitude plane that we use below to seek symbiotic star candidates (\S\ref{sec:machine}). It is defined with these criteria:
\begin{gather}\label{eq:selzone}
    1  \ \text{mag} < \text{\bprp} < 7 \ \text{mag} \\
    \nonumber
    -6 \ \text{mag} < \text{\absmag} < \min(1.8 \times \text{\bprp} - 3. 4)  \ \text{mag},
\end{gather}
where $M_G$ refers to the absolute magnitude in \gaia\ G band. Our goal with these parameter choices is to include a strong majority of known symbiotic stars --- indeed, within this region lie typical red giants \citep[cf.][Fig.~2 therein]{godoy-rivera2025} --- while mitigating contamination by hot supergiants and overwhelming numbers of stars just leaving the main sequence. Within the complete sample set of 248 sources, 221 have GDR3 low-resolution spectra, and of these stars, 171 reside on our selected part of the \CMD. We refer to this subset of 171 stars as our symbiotic star catalog.

In addition to red giants, our catalog contains some noteworthy sources near the perimeter of our selection zone. For example, close to the bright edge is WY Vel, flagged as a variable by Cannon \citep[see][]{shapley1923}. It is a red supergiant with a B-star companion \citep{buss1988, healy2024}, designated as a VV~Cep-type star in the New Online Database of Symbiotic Variables maintained by \citet{merc2019}. Similarly, 17~Lep, just interior to the blue edge of our selected region, is a spectroscopic binary with a red giant and an A-star; mass transfer onto an accretion disk appears to be the result of wind from the red giant \citep{blind2011}. We include these sources to leave open the possibility of identifying new evolved binary systems, noting that location within the selection zone will be important to the ultimate identification of symbiotic stars.

One feature of the \CMD\ in Figure~\ref{fig:systs_cmd} is that sources are not corrected for extinction or reddening. For many stars in our catalog, \gaia\ provides estimates (\gaiavar{ag\_gspphot} and \gaiavar{ebpminrp\_gspphot}), while most others appear in the Bayestar \citep{bayestar2019} and DeCAPS \citep{decaps2025} 3D dust maps, from which we glean that extinction and reddening can be significant. The figure contains a marker showing the location of a star in the \CMD\ and the displacement from reddening of 1~mag. Following \citet{li2023} and \citet{rzhang2023}, we adopt a ratio of G-band extinction ($A_G$) to color excess ratio $E(BP-RP)$ of 3.25. As a measuring stick, this marker suggests that a Be star with an unobscured absolute magnitude $M_G = -3$ and color index $BP-RP = 0$ \citep[e.g.,][Table 5 therein]{radley2025} must be extinguished by about 3.25~mag to reach the faint-bluer tip of our selection zone. Fainter Be stars would miss the selection zone altogether. Wolf-Rayet stars can be even brighter and bluer. With characteristic values of $M_G = -5$~mag and $BP-RP = -0.5$ \citep[cf.][]{mulato2025}, these object require $A_G \approx 5$~mag just to enter our selection zone. Dust extinction at these levels would imply that sources are at low Galactic latitude \citep[e.g.][]{bayestar2019}.

\begin{figure*} 
    \centering
    \includegraphics[width=\figwiddbl]{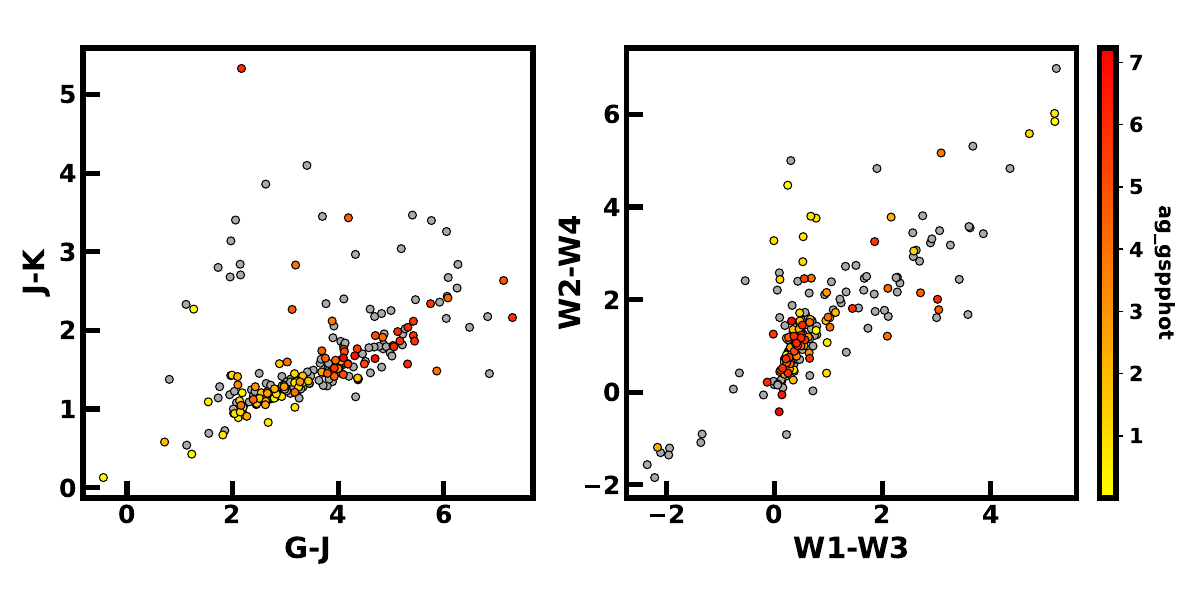}
    \caption{Color-color diagrams of known symbiotic stars in our catalog of 248 sources. In the left plot, J and K magnitudes are from SIMBAD, while the G-band magnitudes are from the \photgmeanmag\ column in the GDR3 main source catalog. On the right are WISE colors. The symbol color of each point designates the extinction (\gaiavar{ag\_gspphot}), provided in the \gaia\ archive, unless it is not provided (gray points).
    \label{fig:systs_colorcolors}}
\end{figure*}

Figure~\ref{fig:systs_colorcolors} shows some evidence for reddening and exctinction in a G-J and J-K color-color diagram for \gaia\ G band, with J and K bands from SIMBAD. Using the full 248-star sample, with 226 stars that have NIR data, we find a discernible correlation between these color indices and the reported extinction in GDR3, indicating mid-infrared absorption and dust, either interstellar or intrinsic to the stellar systems. The figure also shows colors derived from the Wide-field Infrared Survey Explorer \citep[WISE;][]{wise2010, allwise2013}, W1-W3 and W2-W4, to probe intrinsic dust emission. A connection with \gaiavar{ag\_gspphot} is less clear. These mixed results suggest that the extinction and reddening are the result of dust in an indistinguishable mix of interstellar grains along the line of sight (as estimated by 3D dust maps) and dust produced by red giant winds, leading to enrichment of the local environment.

When interstellar extinction is small, color-color relations distinguish distinct types of symbiotic stars \citep{webster1975}. S-type stars have comparatively sparse amounts of dust in the surrounding environment. Their optical and near-infrared (NIR) colors reflect the cool giant's atmosphere with typical effective temperatures of a few thousand Kelvin. D-type symbiotic stars harbor a Mira variable, a late-stage red giant that produces enough dust to impact NIR flux through absorbed and reprocessed starlight. The temperature of the dust in D-type stars is typically below a thousand Kelvin. A third class, D$^\prime$-type stars, have a yellow (perhaps younger) giant \citep[e.g.][and references therein]{akras2019a}. 

Figure~\ref{fig:systs_colorcolors} illustrates that the symbiotic stars considered here have a range of colors in optical and NIR with a clustering in J-K between 1 to 2~mag, typical of red giants. Approximately a fifth of the stars show a NIR excess, with J-K above 2~mag. Interpreting this excess as a sign of the presence of dust, we note that the fraction with $\text{J-K} > 2$  (19\%) is comparable to the fraction of D-type stars in the total symbiotic star population \citep[$\sim$20{\%}][]{belczynski2000, akras2019a}. WISE W1-W3 and W2-W4 colors yield a similar picture. The majority of  sources are  clustered near $\text{W1-W3} \approx 0.5$ and $\text{W2-W4} \approx  1$, presumably representing S-type stars. About 10\%\ of the sources are redder than 2~mag in J-K, with $\text{W1-W3} > 1$ and $\text{W2-W4} > 2$, as expected for D-type stars. 

As a check on our interpretation, we compared our population with the census by \citet{akras2019a} that includes symbiotic star type. The subset of our 248 stars that shows both J-K excess and WISE color excesses, totaling 24 sources, includes 14 D-type stars as identified in \citet[Tables 6 and 7]{akras2019a}. The 174 stars that show no excess are exclusively S-type stars. Overall, the comparison confirms the trend, if not the details, in  relationship between symbiotic star type and the NIR excess reported here.

Clearly, dust plays a prominent role in symbiotic systems, and in many cases its presence is revealed in NIR data. Still, intrinsic reddening and extinction may be indistinguishable from the effect of interstellar dust. For that reason, we choose not to correct photometry for reddening and extinction. This strategy leaves opens the possibility that reddened hot stars may contaminate the sources considered in this work. To cast our wide net, we nonetheless select for a broader range in color-magnitude space (Eq.~(\ref{eq:selzone})) to admit candidates that may be reddened or extinguished, whatever the reason.

\subsection{Astrometric considerations}

The identification of symbiotic stars is facilitated by accurate astrometry, a requirement for precise absolute magnitude calculations. Yet symbiotic stars present a potential challenge. As binaries with the center of visible light tied to the orbiting bright giant, these objects have orbital motion that impacts astrometric measurement. The effect is most acute when the orbital period is comparable to the time baseline of the astrometric observations.  Symbiotic stars consist of roughly solar-mass partners with separations of a few astronomical units, and their orbital periods are that of a couple of years, remarkably similar to the 34-month baseline of GDR3 \citep{gaiadr32023}. Binaries with closer separations and more rapid binary motion would not be resolved by \gaia, while more widely separated pairs have slower orbital motion that does not affect the mutual linear drift that \gaia\ measures.

To assess the impact of double stars on \gaia\ astrometry, we consider the relative change of sky position as a result of orbital motion \citep{stassun2021}. For a binary star with separation of 2~\AU, this motion may shift the sky position of a companion by
\begin{equation}
    {\Delta p} \approx 2 \times \left[\frac{p_0}{1~\displaystyle{\text{mas}}}\right] \ \text{mas},
\end{equation}
where $p_0$ is the parallax in the absence of orbital motion. This shift is roughly equivalent to $p_0$, and is in principle, measurable so long as parallax errors are comparatively low. Adopting a value of $O(0.1)$~mas for the astrometric errors generally, we expect that orbital motion binary stars with $p_0 \gtrsim 0.1$~mas may negatively impact astrometric fits to linear drift, if the orbital period is comparable to GDR3's $\sim$3 year time baseline (for example, Penderghast \textit{et al.}, in preparation). The measure of quality of astrometric fitting in GDR3, \ruwe, is a key indicator of potential double stars; low values indicated good fits, while values above 1.3--1.4 have been specifically associated with orbital motion \citep[see also \citealt{merc2025} for the case of symbiotic stars]{stassun2021, kervella2022, pearce2022, penoyre2022, whiting2023, castroginard2024}. 

\begin{figure}
    \centering
    \includegraphics[width=\figwidsolo]{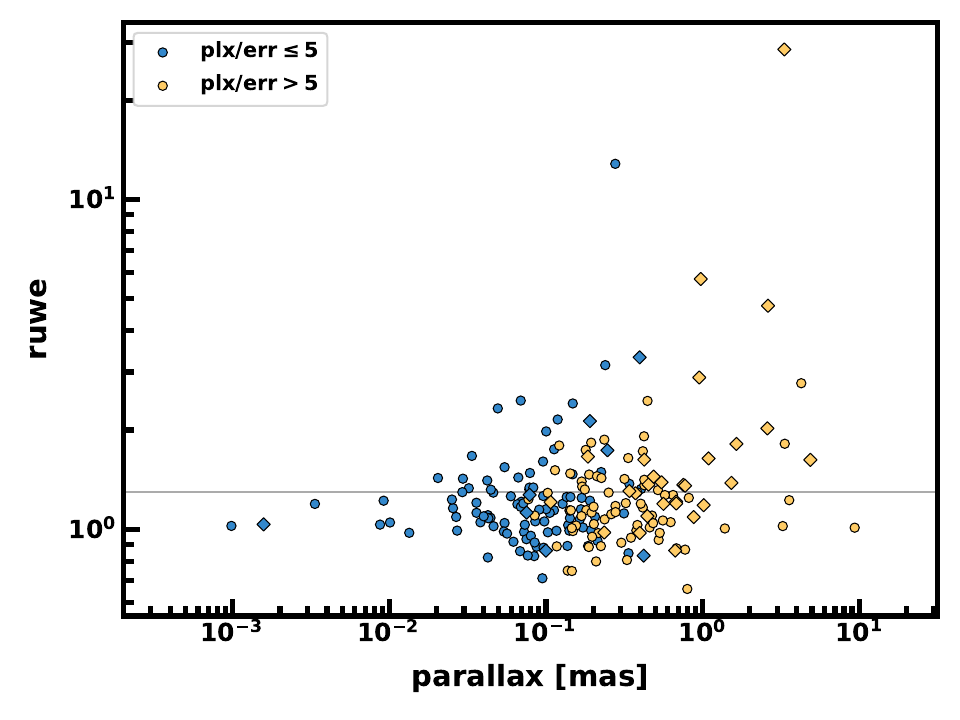}
    \caption{The GDR3 astrometric error parameter \ruwe\ as a function of parallax. The data is consistent with binary stars at a separation of a couple of astronomical units. Nearby sources have higher overall astrometric errors than distant stars, since the orbital displacement on the plane of the sky is greater. If the orbital separation were much smaller or much larger than a few astronomical units, then orbital motion would not have as much of an impact on \gaia\ astrometry, with its 34~month baseline. In the former case, orbital periods would be very short and angular displacement of the binary stars would be small. In the latter case, orbital separations would be large, and binary motion would be too small to be measured astrometrically. As in Figure~\ref{fig:systs_cmd}, X-ray cross-matches are indicated with a diamond symbol, while others are shown as circles. The distribution of X-ray objects is concentrated toward low distances.}
    \label{fig:systs_ruwe_v_parallax}
\end{figure}

Figure~\ref{fig:systs_ruwe_v_parallax}, provides some evidence of this effect in our catalog of symbiotic stars. The majority of the dozen stars with parallax above 1~mas, where orbital motion is clearly resolvable, have \ruwe\ well above 1.3, the nominal threshold for binarity adopted here. Stars with parallax below about 0.1~mas have predominantly low \ruwe, since orbital motion is lost in the noise. In between, stars show a mix of high and low \ruwe, though the majority trend to lower values. We conclude that \ruwe\ is a potentially useful indicator of symbiotic stars, though not a strong one \citep[see also][Appendix A therein]{merc2025}. 

Figure~\ref{fig:systs_ruwe_v_parallax} distinguishes sources in two ways. First, the figure shows X-ray cross-matches (diamond-shape symbols) for comparison with sources with no detected X-ray emission. The X-ray bright sources are concentrated at distances within a few kiloparsecs, the result of flux limits in the survey \citep[e.g.,][]{freund2024}. Second, Figure~\ref{fig:systs_ruwe_v_parallax} indicates sources that have accurate parallaxes ($\text{\poe} > 5$) from those that do not. As expected, more distant sources tend to have lower signal-to-noise in parallax. In light of symbiotic binarity, we might expect to see some close-by objects have high \ruwe\ but also low \poe, since parallax error folds into the general astrometry errors that \ruwe\ characterizes. Indeed some of this trend is apparent but for sources within about 10~kpc, \ruwe\ can still be above 1.3 even when \poe\ is above 5. This observation gives some hope that \ruwe\ may be a valuable indicator for symbiotic stars even when their heliocentric distances are well-measured. 

\subsection{\gaia\ low-resolution spectra and projection coefficients}\label{sec:spec}

Finally, we come to the low-resolution spectra in BP and RP bands provided in GDR3 \citep{deangeli2023, montegriffo2023}. We select for true values of \xpcontinuous\ to facilitate comparison spectra from source to source (data are represented with a common set of template functions and easily yield fluxes at a specified set of spectral wavelengths). We adopt 330 logarthmically-spaced sample points spanning both wavebands over a wavelength range of 361~nm to 1050~nm. We realize each spectrum on this wavelength grid, integrate with Simpson's rule to find the total flux, then normalize so that the total flux is unity in the units set by the \gaia\ BP\_RP spectrum module \texttt{gaiaxpy}. We next subtract off a continuum, identified with a Savitsky-Golay smoothing filter (83-point window width and cubic fitting). To highlight distinctive spectral features as opposed to their strength, we then scale these high-pass filtered signals so that their \textit{pointwise} standard deviation is unity.

\begin{figure}
    \centering
    \includegraphics[width=\figwidsolo]{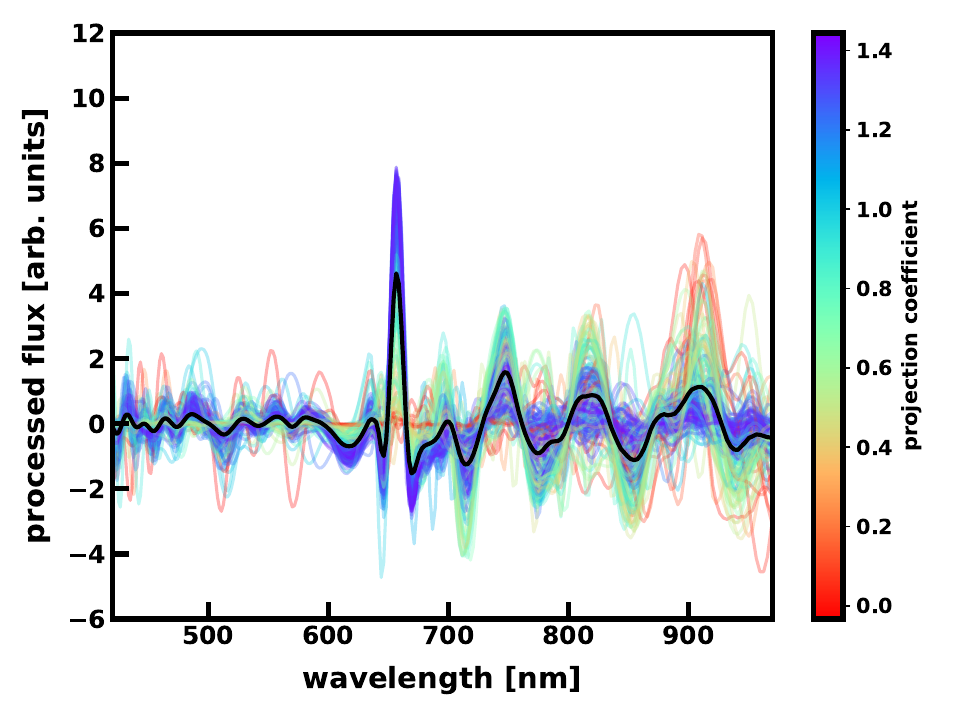}
    \caption{\gaia BP\_RP spectra of sources in our symbiotic star catalog. The thick black curve is the mean spectrum for the catalog, while the thin curves are individual spectra. Their color indicates a projection coefficient, defined in Equation~(\ref{eq:projco}).}
    \label{fig:systs_meanspec}
\end{figure}

Figure~\ref{fig:systs_meanspec} shows the spectra constructed in this way for all 171 objects in our catalog. It indicates that a solid group of sources have comparatively strong \Halpha\ emission and TiO absorption bands, with detectable He~II, \Hbeta, and \Hgamma\ emission lines. Additional emission features that emerge from individual sources include the H Paschen series, O~I, O~II, and [O~III]. Broad emission at 683~nm and 709~nm, characteristic of symbiotic stars \citep{allen1984, kenyon1986} and later identified as Raman scattered O~VI $\lambda\lambda$1032, 1038 emission lines \citep{schmid1989, birriel2000, lee2016, heo2021}, may also be perceptible in some individual spectra. Others have weaker line emission and stronger molecular absorption. Despite these variations, the mean spectrum is a fair representation of individual spectra.

Toward being able to use spectral data to identify symbiotic stars, we quantify the degree to which an individual spectrum matches the mean with a projection coefficient,
\begin{equation}\label{eq:projco}
    \projco_i \equiv \beta(f_i,\bar{f}) = \sum_{j} f_{ij} \bar{f}_j,
\end{equation}
where $f_{ij}$ is the flux from the $i^\text{th}$ star sampled at the $j^\text{th}$ wavelength in the logarithmically spaced set of points, here chosen to run from 600~nm to 775~nm to focus in on \Halpha\ and some of the Ti0 bands, giving both roughly equal weighting. The angular braces denote the catalog average. All spectra are normalized so that $\projco(f_i,f_i) = \projco(\bar{f},\bar{f}) = 1$. The values of $\projco$ range from about -0.03 to 1.44, with an average of unity (by construction) and a median value of 1.18, and a clustering around $S \approx 1.3$. We illustrate this distribution next.  

\subsection{Drawing sources from GDR3}\label{sec:draw}

\begin{figure}
    \centering
    \includegraphics[width=\figwidsolo]{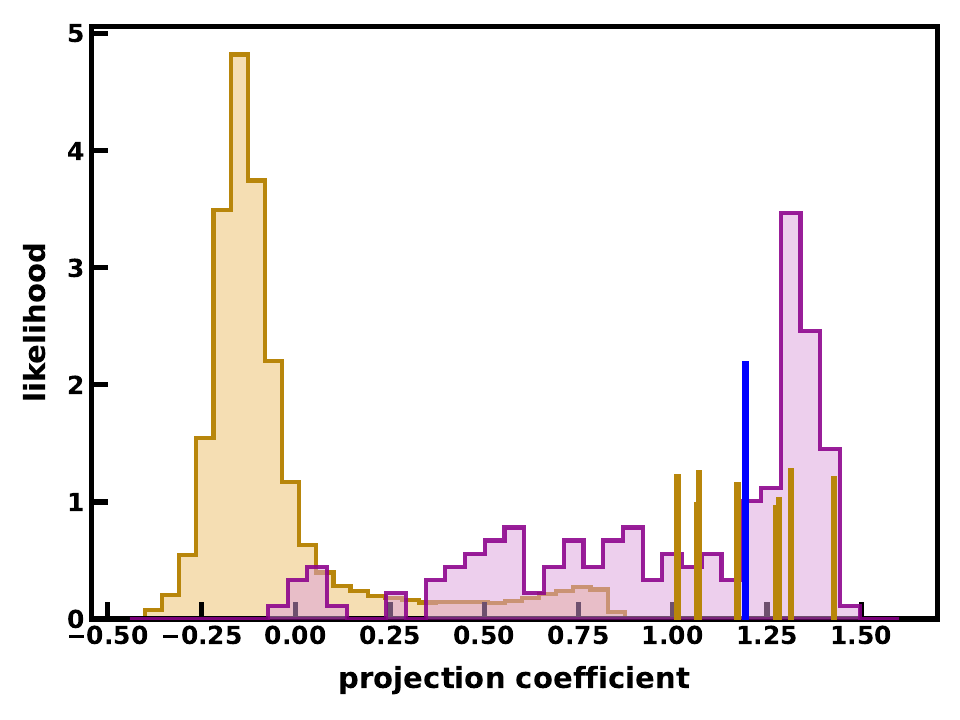}
    \caption{Distribution of projection coefficient $\projco$ from a random draw of over 47,000 sources from \gaia\ (yellow-tinted histogram) and from our sample of symbiotic stars (violet-tinted histogram). Sources with $\projco \gtrsim 0$ corresponds to potential symbiotic star candidates in the random draw, but values above $\projco = 1$ have spectra that more strongly couple to the template function based on the average spectra of known symbiotic stars. The nine sources in the random draw with $\projco > 1$ are indicated with thin, golden-brown vertical lines, including a taller one, in blue, that corresponds to a known symbiotic star that happened to be in the random draw.}
    \label{fig:projco}
\end{figure}

For a preliminary assessment of the projection coefficient $\projco$ as an indicator of symbiotic stars, we perform a random draw from the GDR3 archive from the first ten million sources as ranked by the parameter \gaiavar{random\_index}. This draw netted 47,693 sources, hereafter ``47k stars,'' all lying within our selection zone (Eq.~\ref{eq:selzone}),  brighter than 17~mag in \gaia\ G band, with \poe~$\geq 5$, and with BP\_RP spectral data (the \xpcontinuous\ flag is true).  Figure~\ref{fig:projco} shows a histogram of the projection coefficients of the 47k stars, which are expected to include no more than a few symbiotic stars. The distinction between these sources and symbiotic stars is clear. The vast majority (97\%) of the random stars have $\projco$ values below 0.5, and all but nine lie below $\projco = 1$. Only five of the random stars lie above the median value of $\projco = 1.18$ for symbiotic stars. 
While these outliers may include reddened hot stars as well as symbiotic stars, we conclude that $\projco$ can serve as a powerful discriminator for identifying new symbiotic star candidates.

This example, based on a query of ten million GDR3 sources, represents about 0.6\%\ of the GDR3 archive. Of the nine stars in this group with $\projco > 1$, one, EM*~AS~281, is a known symbiotic source. The other eight sources include promising candidates, one of which, EM AS 276, is listed in SIMBAD as a long-period variable and is discussed further in \S\ref{subsec:promising}. At this rate, we anticipate identifying $\sim$1,500 new symbiotic star candidates in a search of the full archive.

A higher projection coefficient threshold would yield greater selectivity. For example, adopting $\projco \geq 1.18$ (the median value for symbiotics) would lead to under a thousand  candidates. Alternatively, we can use a machine learning algorithm to perform a more nuanced search that includes specific color, magnitude, and astrometric information. We describe our ultimate approach, a combination of both, next.

\section{Seeking symbiotic stars in \GAIA\ DR3} 
\label{sec:method}

With the preliminary exploration based on known symbiotic stars and the sample of 47k stars chosen at random to be in our selected color-magnitude zone (Fig.~\ref{fig:systs_cmd}) and drawn to be fairly bright ($<$17~G mag) and close (by virtue of well-measured parallax, with \poe~$>5$), we train a machine-learning algorithm to identify symbiotic star candidates. The projection coefficients (\S\ref{sec:spec}) will play a significant role, but we will also design the algorithm to take advantage of other information, including astrometry, photometry, and color. We introduce our classifier first, and then proceed to our mining operation to net new symbiotic stars in GDR3.

\subsection{A machine-learning classifier}\label{sec:machine}

To ensure a thorough and well-rounded classifier, we follow this prescription for deriving a machine-learning classifier:
\begin{enumerate}    
\item Set-up training parameters using a select set of columns from the GDR3 main source table to highlight distance, photometry, color, and astrometric solution. Details are below.
\item Download the \gaia\ low-resolution spectra and create projection coefficients for each train/test set. 
\item Create labels to represent if a star is symbiotic or not.
\item Create classifier parameters to fit the datasets and make a matrix regarding the accuracy of the predicted versus misclassified stars.
\item Gather a list of stars from \gaia\ from our selected region outlined in \S\ref{sec:cat}. 
\item Run the classifier, assigning probabilities of a match to symbiotic star properties,  and filter the candidates.
\end{enumerate}

In creating the training set, we use the 171 known symbiotic stars within our selected region. In using this list for the machine learning algorithm (hereafter: ML), we want the algorithm to place extra importance on these specific stars. To make sure this happens, we copied the full 171 over into a new dataset at least once, from there we randomly selected from the 171, allowing duplicates, to add up to a list of 1500 total stars. Each source is assigned a label~$\ell = 1$. We then obtained a sample of non-symbiotic stars from the 47k stars, randomly selecting 1500 stars for our test dataset. The test dataset was then run through SIMBAD to confirm that it contained no known symbiotic stars. These sources were assigned a label~$\ell = 0$. The combined test/train sets totals to 3000 sources. This list was then randomly mixed and split in half (1500 sources) to be tested by the classifier. 

In compiling the spectra data, we calculated the projection coefficients (\S\ref{sec:spec}) of each source and added the column to our test/train datasets for the ML to use (\cof, also \projco). We set up additional \gaia\ attributes on which to further train the data sets. Specifically, we use \bprp, \photgmeanmag, \parallax, \poe, \agof, \ra, \dec, \pmra, \pmdec, \ruwe, and \absmag \ in creating our classifier. The significance of each parameter's inclusion into our model is outlined as follows. 

\begin{itemize}

\item \bprp\ (Blue - Red Color Index): This parameter represents the difference between the BP (blue photometry) and RP (red photometry) magnitudes in the \gaia\ survey. It serves as a proxy for the intrinsic color of a star, which correlates with stellar temperature and spectral type. The color index is critical for distinguishing between stellar populations and classifying stars by their evolutionary stage. Our prior assumptions on \bprp\ are built into selection zone (Eq.~\ref{eq:selzone}) in the color-magnitude plane.

\item \photgmeanmag\ (Mean \gaia~G-band Magnitude): The mean brightness of an object in Gaia’s broad G-band filter, this parameter is fundamental for assessing the apparent luminosity of a source. When combined with parallax, it enables the derivation of absolute magnitudes, which are crucial for distinguishing between different types of celestial objects, such as main-sequence stars, giants, and white dwarfs.

\item \parallax: This parameter provides direct distance measurements to celestial objects.

\item \poe: The signal-to-noise ratio of the parallax measurement, this parameter quantifies the reliability of the distance estimation.

\item \agof\ (Astrometric Goodness-of-Fit Along-Scan): This parameter measures the goodness-of-fit of the astrometric solution in the along-scan direction. It is used to assess the quality of Gaia’s astrometric measurements, with larger values potentially indicating binarity, extended sources, or systematic errors.

\item \ra\ (Right Ascension): This parameter is the measure of an object's position within the sky, specifically measured east-west in hours along the celestial equator.

\item \dec\ (Declination): This parameter is the measure of an object's position within the sky, specifically measured north-south in degrees from the celestial equator.

\item \ruwe\ (Renormalized Unit Weight Error):  This parameter is an indicator of the quality of the astrometric solution. A value significantly greater than unity may indicate binarity when orbital periods are comparable to the \gaia\ time baseline (e.g., \citealt{castroginard2024}; also Penderghast et al., in preparation). The 34-month observations of GDR3 are fortuitiously comparable to the expected orbital period of symbiotic stars, so that \ruwe\ may be a useful indicator in our search (Fig.~\ref{fig:systs_ruwe_v_parallax}).

\item \pmra\ (Proper motion-Right Ascension): This parameter measures the proper motion of an object in the right ascension direction. Proper motion data aides in distinguishing between different stellar populations, such as nearby stars with high proper motion versus distant extragalactic objects with negligible proper motion. This quantity is used to derive the total proper motion, which we adopt as a feature in the machine-leaning algorithm.

\item \pmdec\ (Proper motion-Declination): This parameter measures the proper motion of an object in the declination direction. Like \pmra, it is used in calculating the total proper motion.  


\item \absmag\ (absolute magnitude): The intrinsic brightness of sources derived from \photgmeanmag\ and \parallax\ is a key parameter in the identification of stellar types. This feature-engineering addition is included in part because of its direct role in the definition of the selection zone along with \bprp\ in Eq.~\ref{eq:selzone}.

\item \projco\ (spectral projection coefficient): Since symbiotic stars are ultimately distinguished by the presence of emission lines from hot gas and absorption bands from cool dust, spectroscopy is essential in a search for them. The \gaia\ archive provides spectral information in the form of separate $BP$ and $RP$ coefficients associated with template functions of wavelength. Early investigations used a large subset of these coefficients ($>$40 parameters)  in machine learning, but we performed an extreme dimensional reduction by replacing them with the single projection coefficient in Equation~(\ref{eq:projco}). This quantity encodes the main spectral features that make symbiotic stars unique.

\end{itemize}

\begin{figure}[htbp]
    \centering
    \includegraphics[width=\figwidsolo]{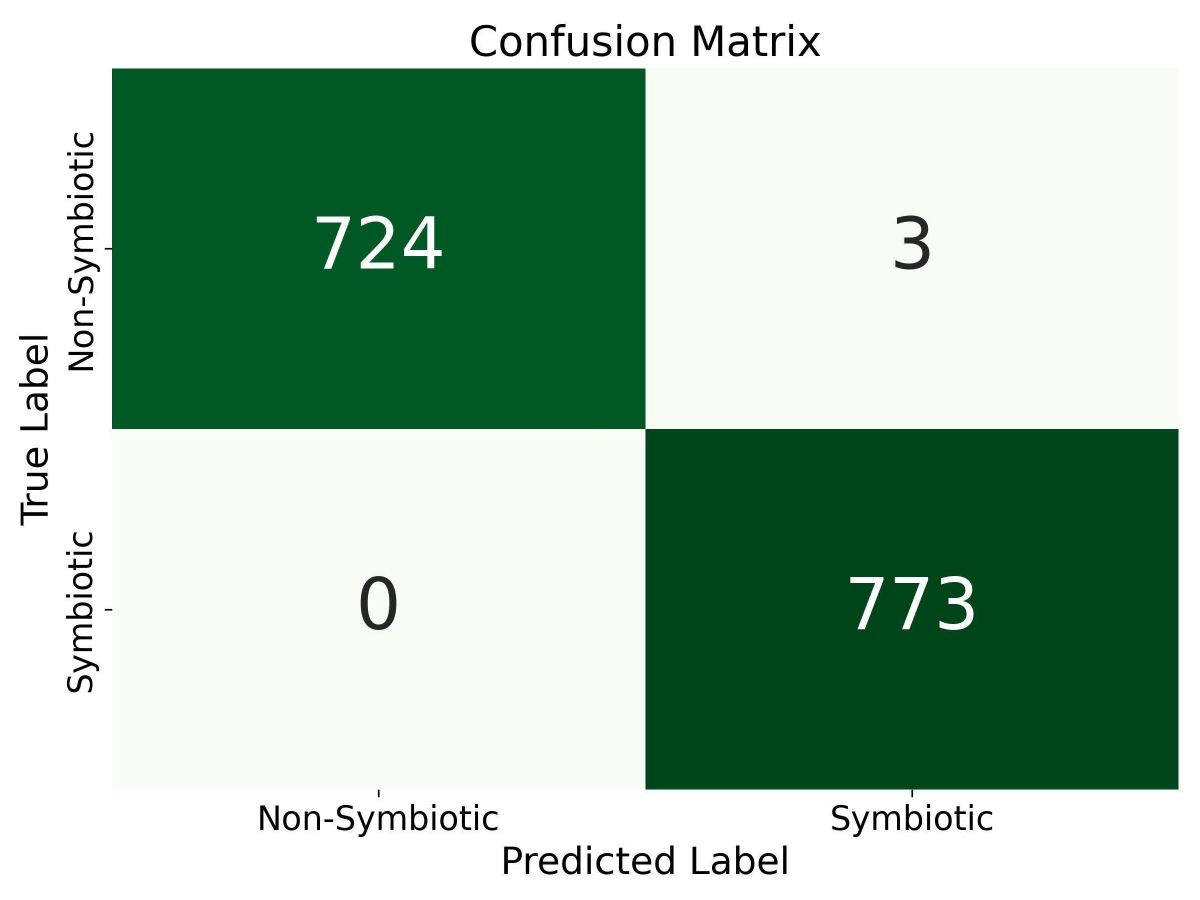} 
    \caption{Confusion matrix of our classifier. The y-axis represents the true labels, with the top row corresponding to true non-symbiotic stars and the bottom row corresponding to true symbiotic stars. The x-axis represents the predicted labels, where the left column indicates what the model predicted as a non-symbiotic star, and the right column indicates what was predicted as a symbiotic star. The upper-left cell (724) represents true negatives, where the model correctly identified the non-symbiotic stars. The upper-right cell (3) represents false positives, where non-symbiotic stars were incorrectly classified as symbiotics. The bottom-left cell (0) represents false negatives, indicating that the model did not miss any true symbiotic stars. The bottom-right cell (773) represents true positives, where all symbiotic stars were correctly identified by the model.}
    \label{fig:matrix}
\end{figure}

We used a parameter grid to fine-tune the Random Forest Classifier for better accuracy. The classifier was then run on the test and training datasets and produced a matrix by which it correctly predicts and/or misclassifies the stars. Figure \ref{fig:matrix} shows the confusion matrix of our classifier, which displays the classification performance of the ML model used on the test dataset to identify symbiotic stars. Overall, the matrix reflects perfect recall (no missed symbiotics) and high precision, with a 1.0 accuracy. It identified only three false positives out of 1,500 predictions—demonstrating the classifier's excellent performance in distinguishing symbiotic stars from the non-symbiotic background population.

\subsection{Mining GDR3 for symbiotic stars}\label{sec:mining}

With a robust machine-learning classifier in place, we select sources from the full GDR3 archive. To come up with a list of candidate symbiotic stars that is manageable, we draw from the archive using the same selection as our 47k stars (eq.~\ref{eq:selzone}, $\text{\photgmeanmag}<17$, $\text{\poe}>5$), and further limit our operation to sources with $\projco >1$, thus focusing on the peak of the projection coefficient distribution for known symbiotic stars. As roughly anticipated from the 47k stars, this selection nets us 1,790 candidates, including 48 known symbiotic stars among the brighter and closer objects in our catalog. 

The next step is to run these sources through the machine-learning classifier.

\section{Results: a new catalog of symbiotic star candidates}\label{sec:results}

Here we present the outcome of applying our classifier (\S\ref{sec:machine}) to the candidate list of stars mined from GDR3 (\S\ref{sec:mining}). The 1,790 candidates, without removing the known symbiotics therein, were run through the ML. Keeping the known symbiotics into our candidate list was a good indicator to see if the classifier would pick out the already known symbiotics. There were 48 known symbiotics in this selection list, the classifier identified the first 30 stars as true symbiotics while the other 18 stars were still within the top 50 respectively. This test demonstrated the accuracy of the classifier. 

After removing the known symbiotics, the ML classifier identified 1,674 stars as candidates with a formal probability $P > 50$\%\ of being a symbiotic star. This list includes 182 candidates with a probability $P$ greater than 90\%, and 36 stars have $P \geq 95$\%.  Table \ref{tab:potentialstars} below shows the 15 stars identified by the classifier with the highest formal probability of a symbiotic star classification. The complete Table~\ref{tab:potentialstars} with all 1,674 candidates is available in this \texttt{arxiv} submission as an ancillary file in machine-readable format (\texttt{table\_2\_all.csv}), along with a separate README file (\texttt{table\_2\_all\_readme.txt}). In Table~\ref{tab:potentialstars}, we indicate whether a source has a cross-match in one of the X-ray catalogs considered here (\rosat, \erosita, and \xmm; see \S\ref{subsec:phot}) to highlight candidates that may have active accretion onto a compact symbiotic partner. 

\begin{deluxetable*}{rrrrrrrrc}
\tabletypesize{\footnotesize}
\tablecaption{Top 15 symbiotic star candidates (ML)
\label{tab:potentialstars}}
\tablehead{
  \colhead{\gaia\ DR3 Source\_ID} &
  \colhead{RA (degrees)} &
  \colhead{Dec (degrees)} &
  \colhead{BP-RP} &
  \colhead{\absmag} &
  \colhead{Ruwe} &
  \colhead{S} &
  \colhead{$p$} &
  \colhead{X?}
}
\startdata 
6908433150198135936 & 309.8358530 & -5.2878887 &   1.875 &  -2.996 & 1.345 &    1.21 &     0.99 \\
5328425487860113792 & 132.9879199 & -48.6426061 &   1.482 &  -1.211 &  3.033 &  1.09 &     0.98 \\
5331436642194644992 & 132.4851801 & -45.4742559 &   1.807 &  -0.130 &  3.442 &   1.11 &     0.97 \\
5258349561694308096 & 151.1261185 & -58.6644456 &   1.389 &  -2.296 & 9.827 &    1.23 &     0.97 \\
3034134257044832000 & 113.3396104 & -11.8110711 &   1.267 &   0.235 & 3.792 &    1.17 &     0.97 \\
5862963206161616640 & 191.9614122 & -62.9959058 &   1.113 &  -2.863 &  1.903 &   1.11 &     0.97 & * \\
5326692730551766016 & 136.8472801 & -48.8349290 &   1.978 &   1.771 &  3.269 &   1.20 &     0.96 \\
5338064773206447744 & 165.4309219 & -60.5850538 &   1.266 &  -1.330 & 2.803 &    1.21 &     0.96 \\
5350817871109520128 & 160.0553395 & -57.8165306 &   1.504 &  -0.123 &  2.702 &   1.11 &     0.96 \\
5983723702088571392 & 242.4964422 & -48.5744313 &   1.641 &  -2.797 & 1.374 &   1.13 &     0.96 \\
6054125191134238336 & 183.0719245 & -63.8280692 &   1.402 &  -0.894 &  1.945 &   1.21 &     0.96 \\
4067633238255477248 & 268.3685113 & -24.7743574 &   1.088 &  -2.883 &  3.636 &   1.28 &     0.96 & * \\
4155000844481174656 & 278.3656995 & -10.5901248 &   1.988 &   1.879 &  8.233 &   1.17 &     0.96  & * \\
5528044641382352640 & 126.6564813 & -39.7441306 &   1.564 &   0.962 & 5.276 &     1.07 &     0.96 \\
512813086494390272 & 21.8378034 & 64.2855732 &   1.514 &  -0.377 &  2.896 &   1.24 &     0.96
\enddata
\tablecomments{Column \absmag\ refers to \gaia\ G absolute magnitude, column \projco\ is the projection coefficient in Eq.~(\ref{eq:projco}), column $p$ is the classifier probability, and in the column labeled $X?$, an asterisk (*) indicates that a source is associated with X-ray emission.}
\end{deluxetable*}

\begin{figure}
    \centering
    \includegraphics[width=\figwidsolo]{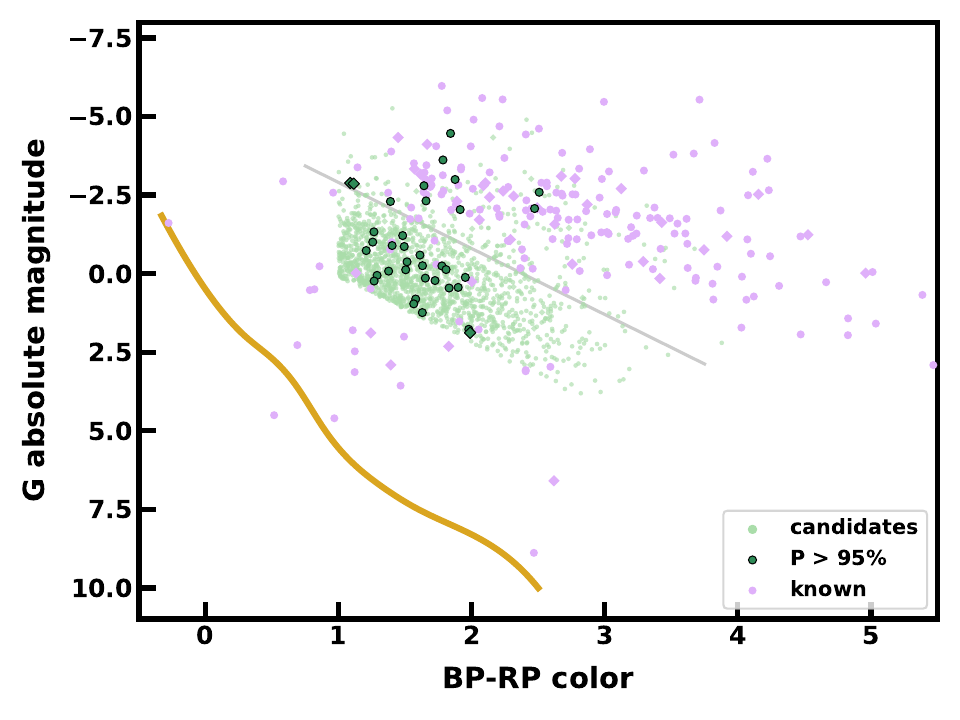}
    \caption{A \gaia-based color-magnitude diagram for candidate symbiotic stars, similar to Figure~\ref{fig:systs_cmd}. Here, the green-toned symbols show our 1,674 candidates, with those have a formal probability of being a symbiotic star at 95\%\ or greater appear in the darker shade. The faint violet-toned symbols are the known symbiotic stars. As in the earlier figure, the orange curve represents the main sequence. A gray line illustrates our boundary to hone in on``red giants,'' sources that are brighter and redder than points on that line yet still within our selection zone. This choice helps identify candidates that are close to the cluster of known symbiotic stars around \absmag\ of $-2$ and \bprp\ of 2.}
    \label{fig:systs_new_cmd}
\end{figure}

To illustrate properties of stars in our catalog of candidates, we first plot a \CMD\ in Figure~\ref{fig:systs_new_cmd}. The distribution of sources in the color-magnitude plane generally increases toward the fainter and bluer region of our selection zone in the \CMD. These objects lie well off the main sequence, but still follow the overall pattern of \gaia\ sources in this part of the \CMD, with the number density increasing closer to the main-sequence ridge line. The qualitative impression is that the bulk of our candidates are at least partly reflective of the underlying \gaia\ source distribution.

Figure~\ref{fig:systs_new_cmd} also shows known symbiotic stars, which more broadly span the selection zone and are loosely clustered around $\text{\bprp} \sim 2$ and $\text{\absmag} \sim -2$, typical of red giants. To highlight these clustered stars, we use a threshold
\begin{equation} \label{eq:redgiant}
    \text{\absmag} < 2.1*\text{\bprp} - 5 \ ,
\end{equation}
indicated with a gray line in the figure. Of the known symbiotic stars, over 90\% are in this red giant region of the \CMD, a measure that guided our choice in adopting Equation~(\ref{eq:redgiant}). In contrast, only 10\%\ of our candidates lie in this red giant zone. For candidates with a 95\%\ probability of being a symbiotic star ($P \geq 0.95)$, this fraction rises to 30\% --- these sources are indicated in Figure~\ref{fig:systs_new_cmd} with dark green symbols. While the ML classifier does not require sources to lie in the zone defined by Equation~(\ref{eq:redgiant}), it evidently recognizes the importance of cool giants to the identification of symbiotic stars. 

\begin{figure}
    \centering
    \includegraphics[width=\figwidsolo]{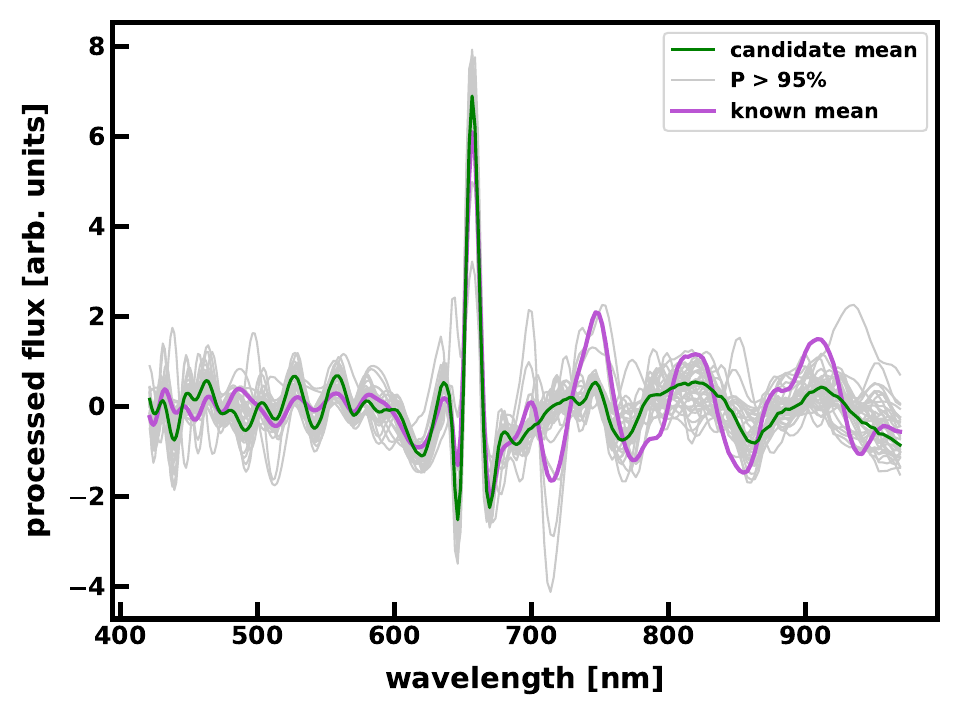}
    \caption{GDR BP+RP spectra of symbiotic star candidates. The thick purple curve is the mean spectrum of known symbiotic stars, the green curve is the average spectrum of all candidate stars, while the thin curves are individual spectra.}
    \label{fig:systs_new_meanspec}
\end{figure}

We selected sources not just on position in the \CMD\ but on a requirement that their \gaia\ spectrum match our template (Fig.~\ref{fig:systs_meanspec}). Figure~\ref{fig:systs_new_meanspec} illustrates, showing the mean low-resolution spectrum of stars in our catalog, including individual spectra of the highest probability sources ($P \ge$~95\%). The basic spectral features include significant H$_\alpha$ emission and possible molecular absorption, with some variability in the strength of these features, as seen with the individual spectra of our top candidates. 

\begin{figure*}[t!]
    \centering
    \includegraphics[width=\figwiddbl]{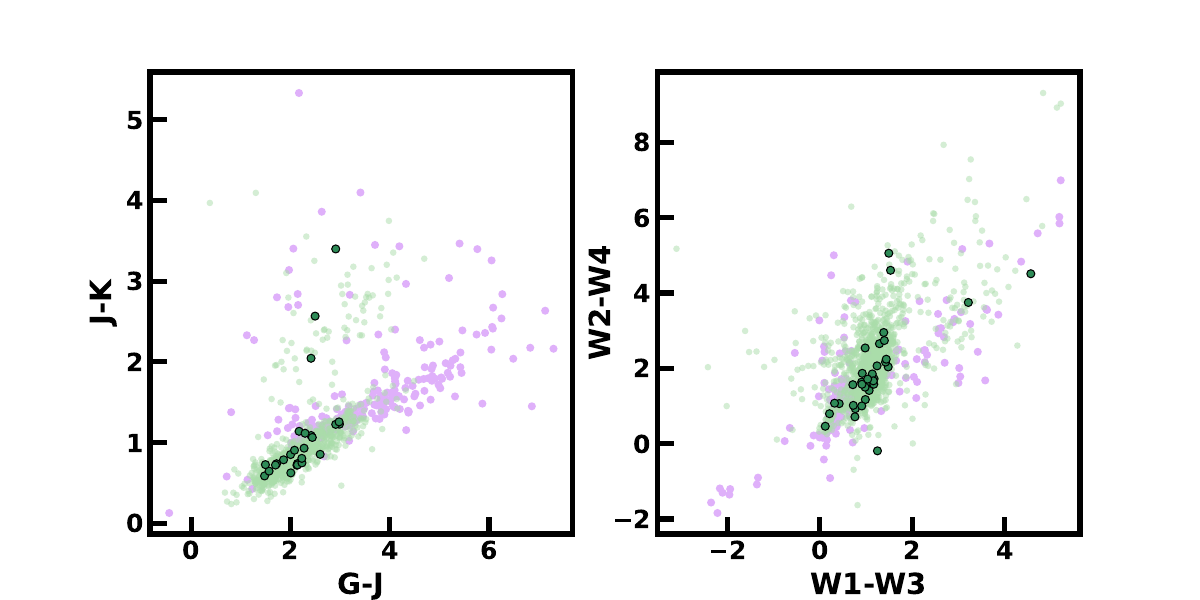}
    \caption{Optical, near infrared and WISE colors for symbiotic star candidates, similar to Figure~\ref{fig:systs_colorcolors}. The candidates are distributed in roughly the same way in these color-color diagrams as known symbiotic stars.}
    \label{fig:systs_new_colorcolors}
\end{figure*}

As an independent check on our candidates' properties, we show available optical, NIR, and WISE color indices of our candidates in Figure~\ref{fig:systs_new_colorcolors}, similar to (cf.~Fig.~\ref{fig:systs_colorcolors}). The overlap in the color-color plane with known symbiotic stars is strong. The candidates lie in a more restricted range of G-J and J-K values, and are on average slightly ($\lesssim 0.5$~mag) bluer than the known symbiotic stars, similar to the distribution in the \CMD\ (Fig.~\ref{fig:systs_new_cmd}). Our candidates' WISE colors also overlap with those of known symbiotic stars. There is a strong clustering in both populations around 0.5-1~mag in W1-W3, and 1-2~mag in W2-W4, with our candidates on average slightly redder than the known symbiotic star population. We interpret these clusters in optical, NIR, and WISE colors as emission from an unobscured giant, as in S-type symbiotic stars. 

Figure~\ref{fig:systs_new_colorcolors} shows color-color outliers among our candidates. These sources are up to several magnitudes redder than the bulk of the stars considered here. Our interpretation, as with the known symbiotic stars, is that these outliers are in a dusty environment, like D-type symbiotic stars.

We emphasize that our candidate selection is entirely determined with \gaia\ data, and that optical-NIR-WISE colors offer an independent avenue for comparisons between candidates and known symbiotic stars.

\begin{figure}
    \centering
    \includegraphics[width=\figwidsolo]{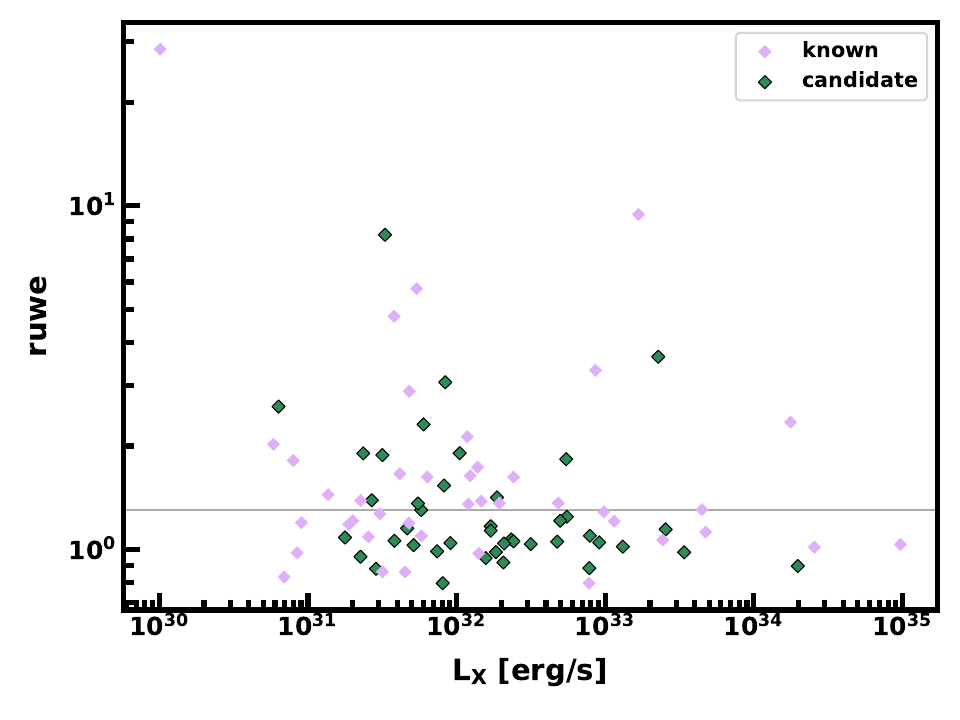}
    \caption{\gaia's astrometric measure \ruwe\ versus X-ray luminosity of candidate symbiotic stars. Candidates introduced in this work are in green, while the light purple symbols indicate known symbiotic stars.}
    \label{fig:systs_new_ruwe_v_xray}
\end{figure}

Finally, to acknowledge the potential importance of X-ray emission and binarity to the identification of symbiotic stars, in Figure \ref{fig:systs_new_ruwe_v_xray} we plot X-ray luminosity and astrometric uncertainty (\ruwe). 
As in \S\ref{sec:method}, we identify potential sources from \rosat, \erosita, and \xmm\ wide-field observations. We derive X-ray luminosities, $L_X$, from the reported X-ray fluxes in each catalog and the geometric distances of \citet{bailerjones2021}. Because of differences between the waveband sensitivity of the X-ray observatories (\rosat\ covers 0.2-2~keV, \erosita\ 0.2-10~keV, and \xmm\ spans 0.1 to 12 keV), as well time variability and X-ray absorption by intervening gas \citep[e.g.,][]{sturm2011}, our inferred X-ray luminosities are order-of-magnitude estimates only. The spread of \ruwe\ values is broad (compare with Fig.~\ref{fig:systs_ruwe_v_parallax}), with a third of our X-ray emitting candidates showing \ruwe\ greater than 1.3, a marker of binarity with separations of a few astronomical units, as expected for symbiotic stars.

Overall, our candidates individually have attributes that make them compelling as potential symbiotic stars. Yet, on a population level, we expect that other stellar types contribute to the demographics. The majority of our candidates are bluer and fainter than typical red giants, and enter our list as a result of the wide net we cast in our search of the Gaia archive. This region of the \CMD\ below the red-giant zone boundary in Figure~\ref{fig:systs_new_cmd}, which we search because known symbiotic stars reside within it, contains comparatively numerous subgiants nearer the main-sequence \citep[cf.][Fig.~30]{gaiacatval2023}. It also emcompasses reddened Be- and O-type stars, which have, like all our candidates, strong \Halpha\ emission. A diagnostic that might inform the possibility of ``contamination'' by these shorter-lived stars is the distance $|z|$ from the Galactic plane, as stars tend to form in the $z=0$ plane and diffuse to greater heights slowly compared with their life times \citep[e.g.,][]{gilmore1989, ting2019, garzon2024, imig2025}. Many of our candidates are at low altitude; 60\% are below $|z| = 100$~pc, comparable to the scale height observed for young, massive stars \citep{bobylev2016, quintana2025}. Others are further from the disk plane. Known symbiotic stars have a broad range of heights, including some at low $|z|$, so we do not exclude sources close to the plane of the Galaxy, focusing instead on other attributes to identify compelling symbiotic star candidates.
    
\subsection{Sources with high promise}\label{subsec:promising}

With the properties of symbiotic stars explored as in \S\ref{sec:method}, these findings highlight the effectiveness of our classifier in isolating potential symbiotic systems from the GDR3 dataset and mark a significant step toward expanding the known catalog of symbiotic stars. Further spectroscopic follow-up is necessary to confirm their symbiotic nature and refine the classification model. 

We can nonetheless hone in on sources with high promise as symbiotic star candidates. We begin by selecting candidates from our catalog of 1,674 stars that have an 80\%\ probability or better of being a symbiotic star. Of these, we consider only those with X-ray emission associated with them. The result is a group of 25 high-promise stars, listed in Table \ref{tab:finalcand}. The table includes \ruwe\ and $L_X$, both of which can bolster the case for interpreting these sources as symbiotic stars along with other spectral and photometric attributes.

\begin{deluxetable*}{rrrrrrrrrc}
\tablecaption{Candidates with associated X-ray emission
\label{tab:finalcand}}
\tablehead{
  \colhead{\gaia\ DR3 Source\_ID} &
  \colhead{RA (degrees)} &
  \colhead{Dec (degrees)} &
  \colhead{BP-RP} &
  \colhead{\absmag} &
  \colhead{Ruwe} &
  \colhead{S} &
  \colhead{$p$} &
  \colhead{L$_X$ (erg/s)} &
  \colhead{Red Giant}
}
\startdata
 5862963206161616640 & 191.9614122 & -62.9959058 &   1.113 &  -2.863 &   1.903 &     1.11 &     0.97 & $2.36\times 10^{31}$ & o \\
 4067633238255477248 & 268.3685113 & -24.7743574 &   1.088 &  -2.883 &   3.636 &     1.28 &     0.96 & $2.27\times 10^{33}$ & o \\
 4155000844481174656 & 278.3656995 & -10.5901248 &   1.988 &   1.879 &   8.233 &     1.17 &     0.96 & $3.30\times 10^{31}$ \\
  511373821468521600 & 26.8679812 & 61.8075708 &   1.852 &  -0.038 &   1.906 &     1.18 &     0.94 & $1.05\times 10^{32}$ \\
 5350304330466080768 & 161.4879820 & -59.9453074 &   1.590 &  -0.723 &   3.066 &     1.32 &     0.94 & $8.39\times 10^{31}$ \\
 5533423903245046912 & 123.6864290 & -41.7564811 &   1.493 &  -0.285 &   1.360 &     1.01 &     0.93 & $5.51\times 10^{31}$ \\
 3429108971530640000 & 88.1401270 & 25.7640650 &   1.346 &  -1.664 &   2.311 &     1.24 &     0.93 & $6.00\times 10^{31}$ \\
 4146612395388085376 & 274.6617474 & -13.7789734 &   1.436 &  -0.506 &   1.831 &     1.32 &     0.93 & $5.47\times 10^{32}$ \\
 4256506967717461120 & 280.2504364 & -5.4643725 &   2.447 &  -3.016 &   1.391 &     1.23 &     0.92 & $2.69\times 10^{31}$ & o\\
 5337634962895857792 & 167.7367796 & -60.7061083 &   1.488 &  -1.401 &   1.245 &     1.09 &     0.91 & $5.53\times 10^{32}$ \\
 4097811804449201664 & 275.3833836 & -16.2233348 &   1.559 &   0.860 &   1.535 &     1.24 &     0.90 & $8.24\times 10^{31}$ \\
 2060601411718713088 & 304.8842007 & 37.2875676 &   1.177 &  -1.270 &   2.604 &     1.13 &     0.89 & $6.34\times 10^{30}$ \\
 5866068158280977536 & 210.2027359 & -62.2514152 &   2.314 &   0.420 &   1.169 &     1.08 &     0.89 & $1.70\times 10^{32}$ \\
 4323101294616761472 & 293.4896372 & 18.4139288 &   2.105 &  -0.897 &   1.303 &     1.11 &     0.88 & $5.78\times 10^{31}$ & o \\
 5963997295254312064 & 253.3623541 & -45.4071775 &   1.669 &  -0.268 &   1.041 &     1.23 &     0.85 & $2.09\times 10^{32}$ \\
 4319930096909297664 & 290.3915616 & 14.8824531 &   2.161 &  -4.331 &   0.982 &     1.28 &     0.85 & $1.84\times 10^{32}$ & o \\
  276644757710014976 & 64.9255612 & 55.9993607 &   1.584 &  -2.622 &   1.144 &     1.28 &     0.85 & $2.55\times 10^{33}$ & o\\
 5255677056923564544 & 156.0766151 & -57.8082591 &   2.730 &  -1.456 &   1.048 &     1.01 &     0.84 & $9.12\times 10^{32}$ & o \\
 5866510642950541952 & 213.5103447 & -61.6919817 &   1.327 &  -1.417 &   1.070 &     1.25 &     0.84 & $2.33\times 10^{32}$ \\
 2067537096694188416 & 305.5259549 & 40.4465447 &   2.528 &   1.232 &   1.083 &     1.23 &     0.83 & $1.78\times 10^{31}$ \\
 3441214006141133952 & 84.6512611 & 26.5049398 &   1.348 &   0.039 &   1.152 &     1.03 &     0.82 & $4.65\times 10^{31}$ \\
 5976051206857091200 & 261.1779942 & -34.1958969 &   2.441 &   2.905 &   1.057 &     1.02 &     0.82 & $2.42\times 10^{32}$ \\
 5966221298030125056 & 254.7781823 & -42.7023476 &   2.173 &  -0.531 &   1.882 &     1.26 &     0.81 & $3.17\times 10^{31}$ & o \\
  519352324516039680 & 23.9577173 & 66.2120235 &   2.003 &  -1.427 &   1.054 &     1.29 &     0.81 & $4.75\times 10^{32}$ & o \\
 4153412355116242688 & 277.1774747 & -11.2828971 &   2.469 &   1.747 &   1.096 &     1.02 &     0.81 & $7.89\times 10^{32}$
 \enddata
 \tablecomments{Column L$_X$ is the total X-ray flux in erg/s. The column listed as Red Giant, denoted by a lower case (o), is stars found by Eq \ref{eq:redgiant} indicated above the grey line in our selection zone in Figure \ref{fig:systs_new_cmd}. }
\end{deluxetable*}

Table \ref{tab:finalcand} also includes a column indicating whether a source is associated with a red giant. Our motivation for including this indicator stems from Figures~\ref{fig:systs_cmd} and \ref{fig:systs_new_cmd}. They illustrate that symbiotic stars, while distributed over a range of colors and magnitudes, tend to cluster in the region of the \CMD\ associated with red giants (Eq.~(\ref{eq:redgiant}). This yields nine stars noted in the table by ``o'' in the Red Giant column. 

We further highlight a subset of manually selected candidates from our broader sample of 1,674 stars that were not included in Tables \ref{tab:potentialstars} and \ref{tab:finalcand} but demonstrate intriguing characteristics suggestive of symbiotic behavior. These selections are based on a combination of high \ruwe\ values, strong projection coefficients (S values), and favorable SIMBAD object types (o-types), particularly those identified as emission-line stars (EM*), X-ray sources (X), or long-period variables (LP*). While their ML probabilities may not place them among the highest-ranked candidates, these additional observational features warrant further notation. One such candidate is Gaia DR3 524428838421397760, which exhibits a very high \ruwe\ value of 9.990 and a strong projection coefficient (S value) of 1.24. SIMBAD identifies this source as EM GGA 2*, a long-period variable star. Although it lacks associated X-ray information, being one of the highest ruwe values in our sample makes it a compelling symbiotic candidate.

Gaia DR3 5888269634519939712 presents a similarly strong case with $\projco = 1.43$, one of the highest in our sample, and a \ruwe\ value of 1.090, lower than that of EM*~GGA~2 but still noteworthy. This star is listed in SIMBAD as [M81]~I-468, also categorized as a long-period variable. Like the previous star, it shows no X-ray emission, yet its projection coefficient places it among the more promising candidates.

Lastly, \mbox{Gaia DR3 6725263992006252160} has an unusually high \projco\ value of 1.43. It is identified in SIMBAD as \mbox{EM* AS 276}, currently listed as a long-period variable candidate. However, \citet{allen1978} recognized \mbox{AS 276} (object 135 therein) as a symbiotic star, describing it as having He II weaker than H$\beta$, strong TiO absorption, absence of \mbox{[Fe VII]} and \mbox{[O III]}, and a strong \mbox{[Fe VI]} line with a prominent $\lambda6830$ emission band. Despite the absence of X-ray data, this earlier classification, together with the photometric and spectral characteristics highlighted here, strongly supports its symbiotic nature \citep[see also][]{kenyon1986}. Its recovery by our ML approach, operating solely on \gaia's astrometric, photometric and spectroscopic parameters, demonstrates the method’s ability to identify genuine symbiotic stars.

\section{Conclusion}\label{sec:conclude}

Symbiotic stars offer important windows into stellar and binary evolution. They represent a prominent evolutionary phase as late-type stellar partners leave the main sequence, each star in its turn, to become double white dwarf binaries. These compact systems give rise to Type Ia supernovae \citep{munari1992, distephano2010}, standard candles that are essential in astrophysical cosmology \citep{brout2022}.

Stellar evolution theory and stellar population synthesis tie together our understanding of symbiotic stars and their prevalence in the Milky Way \citep{kenyon1986, mmunari1992, yungelson1995, magrini2003, lu2006, laversveiler2025}. Yet predictions exceed reality by a factor of ten or more. One reason is the cool giant dominates the light at optical wavebands, so symbiotic stars hide among other giant stars in photometric surveys. Another reason is that these stars are variable in their spectral features \citep[e.g.,][]{merc2025review}. Recent surveys \citep[e.g.,][]{akras2019a, akras2019b, akras2021, akras2023, jia2023, xu2024, chen2025} have made progress, but a gap still remains between detections and theoretical predictions. 

Our project stems from a desire to improve this situation. We worked with the \gaia\ Data Release~3 archive, identifying a subset of 171 known symbiotic stars observed with Gaia's low resolution spectrometers. We defined selection criteria in color, magnitude, astrometric quality, and a projection coeffecient ($\projco$ in Eq.~(\ref{eq:projco})) that measures how well-matched a Gaia spectrum is to the characteristic spectrum of a symbiotic star. Noting that other factors may help to identify these stars, including astrometric uncertainties because of binarity, we trained a machine-learning classifier to provide a refined list of candidates from the 1.8 billion Gaia sources.  

The result is a catalog of 1,674 symbiotic star candidates. We recommend follow-up observations for all of these sources. In some cases where we have X-ray counterparts, we are most confident that we have identified new symbiotic stars, and recommend prioritizing these sources for further study.

Our approach balances accuracy in astrometry, brightness of candidates, and broad spectral features to search for new symbiotic candidates.  While our study has limitations, including the brightness cuts and dependence on a single spectral template, future work could improve upon this strategy to cast an even wider and more selective net. The rewards would be significant. With each new discovery comes the potential for a deeper understanding of stellar evolution and the stars that have revolutionized our understanding of the universe.

\vspace*{1pt}
\section*{Acknowledgments} 
We thank an anonymous referee for comments and suggestions that improved our work. SEB is grateful for support from the Undergraduate Research Opportunity Program at the University of Utah. This research has made use of data from the European Space Agency (ESA) mission {\it Gaia} (\url{https://www.cosmos.esa.int/gaia}), processed by the {\it Gaia} Data Processing and Analysis Consortium (DPAC, \url{https://www.cosmos.esa.int/web/gaia/dpac/consortium}). Funding for the DPAC has been provided by national institutions, in particular the institutions participating in the {\it Gaia} Multilateral Agreement. This research has also made use of the SIMBAD database \citep{simbad2000}, and the VizieR catalog access tool \citep{vizier2000}, operated at CDS, Strasbourg, France. We are thankful for the availability of these resources.

\bibliographystyle{aasjournal}
\bibliography{main}{}

\end{document}